\documentclass[aps, prb, twocolumn, showpacs, superscriptaddress]{revtex4-1}
\usepackage{bm,amssymb,amsmath}
\usepackage{times}
\usepackage{graphicx}
\usepackage{color}
\usepackage{dcolumn}
\usepackage[colorlinks=true, letterpaper=true, pdfstartview=FitV, linkcolor=blue, citecolor=blue, urlcolor=blue]{hyperref}
\usepackage[normalem]{ulem}

\begin{document}

\title{Classification of spin Hall effect in two-dimensional systems}
\author{Longjun Xiang}
\affiliation{College of Physics and Optoelectronic Engineering, Shenzhen University, Shenzhen 518060, China}
\author{Fuming Xu}
\affiliation{College of Physics and Optoelectronic Engineering, Shenzhen University, Shenzhen 518060, China}
\affiliation{Quantum Science Center of Guangdong-Hongkong-Macao Greater Bay Area (Guangdong), Shenzhen 518045, China}
\author{Luyang Wang}
\email[]{wangly@szu.edu.cn}
\affiliation{College of Physics and Optoelectronic Engineering, Shenzhen University, Shenzhen 518060, China}
\author{Jian Wang}
\email[]{jianwang@hku.hk}
\affiliation{College of Physics and Optoelectronic Engineering, Shenzhen University, Shenzhen 518060, China}
\affiliation{Department of Physics, The University of Hong Kong, Pokfulam Road, Hong Kong, China}
\begin{abstract}
Physical properties such as the conductivity are usually classified according to the symmetry of the underlying system using Neumann's principle, which gives an upper bound for the number of independent components of the corresponding property tensor. However, for a given Hamiltonian, this global approach usually can not give a definite answer on whether a physical effect such as spin Hall effect (SHE) exists or not. It is found that the parity and types of spin-orbit interactions (SOIs) are good indicators that can further reduce the number of independent components of the spin Hall conductivity for a specific system. In terms of the parity as well as various Rashba-like and Dresselhaus-like SOIs, we propose a local approach to classify SHE in two-dimensional (2D) two-band models, where sufficient conditions for identifying the existence or absence of SHE in all 2D magnetic point groups are presented.
\end{abstract}
\maketitle

\section{introduction}

Berry curvature related band geometric quantities are widely adopted to describe various Hall effects.\cite{D-Xiao,Gao19} For instance, nonzero Berry phase accounts for both quantum Hall effect\cite{QHall} and quantum anomalous Hall effect,\cite{QAHall} which are intrinsic responses and involve the breaking of time-reversal symmetry. In time-reversal invariant systems, extrinsic Hall effect can exist in the nonlinear response regime, such as the second-order Hall effect induced by Berry curvature dipole\cite{L-Fu,Guinea1,Du2021} and the third-order Hall effect induced by Berry-connection polarizability tensor\cite{ThirdHE2021}, which can also be studied in multiterminal systems using the scattering matrix theory.\cite{Wei222,Wei223} Higher-order nonlinear anomalous Hall effects induced by Berry curvature multipoles such as quadrupole and hexapole have been discussed in certain materials with magnetic point group symmetry\cite{Law}. In particular, intrinsic second-order anomalous Hall effect has been discovered in $\mathcal{PT}$-symmetric antiferromagnets\cite{Yang,Xiao,Gao23} and intrinsic third-order anomalous Hall effect was discussed.\cite{Xiang23} In addition to Hall currents, it was found that Berry curvature dipole and Berry curvature fluctuation can give rise to linear and nonlinear thermal Hall noises.\cite{Wei23} The existence of these nonlinear Hall effects are classified by symmetry and relevant constraints\cite{L-Fu,Guinea1,Yang,Xiao,Law,Xiang23}, since band geometry is strongly affected by symmetry.

In fact, symmetry plays essential roles in classifying a variety of physical properties.\cite{Lu23} For example, the famous universal conductance fluctuation (UCF)\cite{Altshuler,Lee1,Lee2,CWJBeenakker97,qiao2019FOP} in mesoscopic transport depends on symmetry and dimensionality of the system. To describe UCF, the system Hamiltonian is categorized into three ensembles, i.e., Gaussian orthogonal/unirary/sympletic ensemble, based on the presence or absence of time-reversal and spin-rotation symmetries. When the particle-hole and chiral symmetries are further included, it has been extended to the ten-fold way\cite{Tenfoldway,Ryu}, which is widely used in classifying topological insulators (TI) and topological superconductors. Spin photovoltaic effect in antiferromagnetic materials can be classified in terms of the $\cal{PT}$ symmetry and spacial symmetries.\cite{Jiang23} However, there are also exceptions using symmetry analysis. The remarkable example is the phase transition between TI and band insulator, which occurs without symmetry breaking but with the changing of global (topological) invariants\cite{XLQi,Bernevig}. In this work, we find another example: for a system with a given symmetry, spin Hall effect (SHE) can be present or absent depending on the parity and spin-orbit interaction (SOI) of the Hamiltonian.

SHE is a relativistic phenomenon, where charge current drives transverse spin current in spin-orbit coupled systems\cite{SHE}. Similar to charge Hall effects, SHE can also be intrinsic or extrinsic. Intrinsic SHE\cite{Sinova2004,Yang11} is not influenced by transport process, which has been proposed in Weyl semimetals such as TaAs\cite{Sun2016}. The orientation of spin polarization in SHE can be either in-plane or out-of-plane. In different materials, the dominant SOI could be Dresselhaus-like or Rashba-like, or the combination of them\cite{Hopfner,Nechaev,Bellaiche}. The existence of SHE and inverse SHE\cite{Sinova1} have been solidly confirmed by a series of experiments\cite{Wunderlich,Tatara}, and theoretical description of SHE usually involves spin Berry curvature\cite{Sun2016}. However, less attention has focused on classifying SHE, which could be carried out using Neumann's principle. Neumann's principle has been applied to analyze the conductivity tensor for nonlinear Hall effects\cite{Law,Yang,Xiao}, which can be stated as: if a crystal is invariant with respect to certain symmetry, any of its physical properties must also be invariant with respect to the same symmetry. As a consequence of this principle, symmetry imposes constraint on physical quantities, including the spin Hall conductivity.

In this work, we demonstrate that the symmetry alone is insufficient to determine the existence of SHE in 2D systems, since SHE can be switched on or off by varying the SOI while maintaining the symmetry of the system. Compared with Neumann's principle, the parity and types of SOIs are better indicators for identifying SHE. Based on the parity as well as different orders of in-plane or out-of-plane, Dresselhaus-like or Rashba-like SOIs, we propose a local approach to classify SHE, which gives sufficient conditions for the existence of SHE in 2D two-band models. Complete analysis on all 2D magnetic point groups (MPGs) are carried out and possible materials for experimental verification are discussed.

The paper is organized as follows. In Sec.~{\color{blue}II}, the Kubo formula for the spin Hall conductivity of a 2D two-band model is introduced. In Sec.~{\color{blue}III}, classification of 2D SHE according to Neumann's principle is discussed. For a Hamiltonian with certain symmetry, this general classification can only give an indefinite answer on the presence or absence of SHE. In Sec.~{\color{blue}IV}, a different symmetry analysis based on parity and types of SOIs constituting the 2D Hamiltonian is given, which allows identifying the existence of SHE. Finally, discussion and conclusion are given in Sec.~{\color{blue}V}.

\section{formalism for spin Hall effect}

The spin Hall conductivity is\cite{Sinova2004,Schliemann,Sun2016} ($\hbar=e=1$)
\begin{eqnarray}
\sigma_{xy}^{\alpha}=\int_k \sum_n f_{n,{\bf k}}\Omega_{n,xy}^\alpha({\bf k}),\label{Kubo}
\end{eqnarray}
where $\int_k =\int_{BZ} d^2k/(2\pi)^2$ and $\alpha$ labels the direction of spin polarization of SHE. The "spin" Berry curvature is defined as
\begin{eqnarray}
\Omega_{n,xy}^\alpha({\bf k})=-2{\rm Im}\sum_{n'\ne n}\frac{\langle n|{J}_x^\alpha|n'\rangle \langle n'|{v}_y|n\rangle} {(\epsilon_{n}-\epsilon_{n'})^2},
\end{eqnarray}
with spin current operator $J_i^\alpha=\frac{1}{2}\{{v}_i,{s}_\alpha\}$, where ${s}_\alpha$ is the spin operator and ${v}_i=\frac{\partial H}{\partial k_i}$ is the velocity operator. Note that the "spin" Berry curvature resembles usual Berry curvature only if the spin is a good quantum number. In the presence of spin-orbit interaction (SOI), it is very different from the usual Berry curvature. We focus on a two-band model defined as $H = d_0\sigma_0 + {\bf d} \cdot {\bm {\sigma}}$,
from which we have
\begin{eqnarray}
v_i=\partial_i d_0+(\partial_i d_j)\sigma_j,
\end{eqnarray}
where summation over repeated indices is implied and
\begin{eqnarray}
J_i^\alpha=\frac{\hbar}{2}(\sigma_\alpha\partial_i d_0+\partial_i d_\alpha).
\end{eqnarray}

\renewcommand\arraystretch{1.2}
\begin{table*}[tbp]
	\begin{center}
		\caption{\label{table1} SOI Hamiltonians of various orders and symmetries. }
		\begin{tabular}{p{9cm}<{\centering}|p{8cm}<{\centering}}
\hline\hline
in-plane Dresselhaus-like SOI (symmetry)    & in-plane Rashba-like SOI (symmetry) \\ \hline			
$H_2 = {\Re} (k_+\sigma_+ )$ ~ (21')  & $H_1 = k_y\sigma_x-k_x\sigma_y \sim {\Im} (k_+\sigma_- )$ ~ ($C_\infty$ , $\cal T$)\\ \hline
$H_{3x} = {\Re}(k_+^2\sigma_-)$  (2'mm')\cite{Nechaev,Acosta}, ~ $H_{4x} ={\Re}(k_+^2\sigma_+)$  (6'm'm)\cite{Nechaev,Acosta}  &  $H_{3y} ={\Im}(k_+^2\sigma_-)$  (2'm'm)\cite{Nechaev}, ~ $H_{4y} ={\Im}(k_+^2\sigma_+)$  (6'm'm)\cite{Nechaev} \\ \hline
$H'_{2} ={\Re}(k_+^3\sigma_-)$  (21'), ~ $H_{5} = {\Re}(k_+^3\sigma_+)$  (41')& $H_{6} ={\Im}(k_+^3\sigma_-)$  (2mm1')\cite{Nechaev}, ~ $H_{7} ={\Im}(k_+^3\sigma_+)$  (4mm1')\cite{Nechaev}\\ \hline
$H'_{5} ={\Re}(k_+^5\sigma_-)$  (41'), ~ $H_{8} = {\Re}(k_+^5\sigma_+)$  (61')& $H'_{7} ={\Im}(k_+^5\sigma_-)$  (4mm1'), ~ $H_{9} ={\Im}(k_+^5\sigma_+)$  (6mm1')\\
\hline \hline
out-of-plane Dresselhaus-like SOI     & out-of-plane Rashba-like SOI \\ \hline	
${\Re}(k_+)\sigma_z$ (m1')\cite{Stolwijk} & ${\Im}(k_+)\sigma_z$ (m1') \\ \hline
${\Re}(k_+^2)\sigma_z$ ~ (4'm'm)\cite{Law} & ${\Im}(k_+^2)\sigma_z$ ~ (4'm'm)\\ \hline
${\Re}(k_+^3)\sigma_z$ ~ (3m)\cite{L-Fu3,Mich,Bellaiche,Vajna}& ${\Im}(k_+^3)\sigma_z$ ~ (3m)\cite{Paul,Bellaiche,Vajna,Bahramy}\\
\hline\hline			
		\end{tabular}
	\end{center}
\end{table*}

After some algebra, we find the spin Berry curvature for the lower band
\begin{eqnarray}
\Omega_{xy}^\alpha({\bf k})=\frac{\partial_x d_0(\partial_y{\bf d}\times{\bf d})_\alpha}{4d^3}, \label{spH}
\end{eqnarray}
where $d^2 = {\bf d} \cdot {\bf d}$. It is easy to show that for the linear Rashba SOI, Eq.~(\ref{spH}) reproduces the spin Hall conductance of $e/8\pi$, which was obtained by Sinova et al\cite{Sinova2004}. For comparison, we show the expression of Berry curvature
\begin{eqnarray}
\Omega_{\pm,xy}({\bf k})=\pm \frac{\hbar\partial_x {\bf d} \cdot (\partial_y{\bf d}\times{\bf d})}{2d^3}. \label{BC}
\end{eqnarray}
Obviously, if $H_0 =0$ there is no linear SHE. From now on, we will work on systems with broken particle-hole symmetry ($d_0 = k^2$) and focus on $\sigma^{\alpha}_{xy}$. The analysis of $\sigma^{\alpha} _{yx}$ is similar.

In order to have spin Hall effect, SOI must be present which can be classified according to the symmetry as well as linear or nonlinear orders of momentum. For instance, two typical in-plane linear order SOIs (IP-SOI) in 2D systems are the Rashba SOI $k_y \sigma_x - k_x \sigma_y$ and Dresselhaus SOI $k_x \sigma_x - k_y \sigma_y$.
For the classification reason, we define two types of SOI, Dresselhaus-like ($H_D$) and Rashba-like ($H_R$) SOI, as follows. For IP-SOI, we define $A = \sum_{n=1}^N [\alpha_n k_+^{N-n} k_-^n \sigma_+ + \beta_n k_-^{N-n} k_+^n \sigma_-]$ with $k_\pm = k_x \pm i k_y$, then $H_D = A + A^\dagger$ and $H_R = -i (A - A^\dagger)$. For the out-of-plane SOI (OP-SOI), we have $H_D = B + B^\dagger$ and $H_R = -i (B - B^\dagger)$ where $B = \sum_{n=1}^N \gamma_n k_+^{N-n} k_-^n \sigma_z$.
In Table \ref{table1}, we list a few IP-SOIs and OP-SOIs which are basic building blocks of the SOI Hamiltonian. In particular, the SOI with $N=3$ is called cubic Dresselhaus and Rashba SOI, respectively in the literatures\cite{Gerchikov, Schliemann, Bleibaum,Moriya, Bellaiche}. According to our notation, $k_y^3\sigma_x - k_x^3 \sigma_y \sim {\rm Im} [(k_+^3 +3 k_+ k_-^2)\sigma_+]$, $(k_x^2 - k_y^2)(k_y\sigma_x + k_x\sigma_y) \sim {\rm Im} [(k_+^2 + k_-^2)k_+\sigma_+]$, $k_x^2 k_y^2 (k_y\sigma_x - k_x\sigma_y) \sim {\rm Im} [(k_+^2 - k_-^2)^2 k_+\sigma_-]$, and $k_x k_y (k_x\sigma_x - k_y\sigma_y) \sim {\rm Im} [(k_+^2 - k_-^2)k_+\sigma_-]$ belong to Rashba-like SOI with $C_{4v}$ symmetry\cite{Shanavas,Arras,Silveira} while $k_x k_y (k_y\sigma_x - k_x\sigma_y) \sim {\rm Re} [(k_+^2 - k_-^2)k_+\sigma_-]$ and $k_x^3\sigma_x + k_y^3\sigma_y \sim {\rm Im} [(k_+^3 +3 k_+ k_-^2)\sigma_+]$ belongs to Dresselhaus-like SOI with $C_4$ symmetry\cite{Marinescu,Vajna}.

\section{Classification of SHE from Neumann's principle}

SHE can be classified using Neumann's principle. Note that the spin current is obtained from spin (pseudo-vector) and velocity (vector) operators, it forms a second rank pseudo-tensor while the electric field is a vector. As a result, the spin conducticity tensor $\sigma^\alpha_{\beta\gamma}$ is a third rank pseudo-tensor ($\beta, \gamma = x, y$).\cite{Glazova,Ebert} Different from the Hall effect, generally speaking, spin conductance does not enjoy antisymmetric property with respective to $\beta$ and $\gamma$. According to Neumann's principle, the spin conductivity tensor is expressed in terms of a rotation matrix $R$ as
\begin{eqnarray}
\sigma^\alpha_{\beta\gamma} = \eta {\rm det}(R) R_{\alpha\alpha'} R_{\beta\beta'} R_{\gamma\gamma'}\sigma^{\alpha'}_{\beta'\gamma'}. \label{trans}
\end{eqnarray}
where $\eta = -1$ for prime operation and the presence of ${\rm det}(R)$ is needed for a pseudo-tensor. From Eq.~(\ref{Kubo}), it can also be shown as
\begin{eqnarray}
\sigma^\alpha_{\beta\gamma}=\int_k \Omega^\alpha_{\beta\gamma}({\bf k})=\eta {\rm det(R)} \int_k \Omega^\alpha_{\beta\gamma}(R{\bf k}), \label{trans2}
\end{eqnarray}

Since 2D MPG is imbedded in 3D MPG, we can use Bilbao Crystallographic Server\cite{Bilbao} to find nonzero components of $\sigma^\alpha_{\beta\gamma}$, which is similar to the discussion of higher-order anomalous Hall effects.\cite{Law,Yang,Xiao}

\begin{figure}
\centering
\includegraphics[width=\columnwidth]{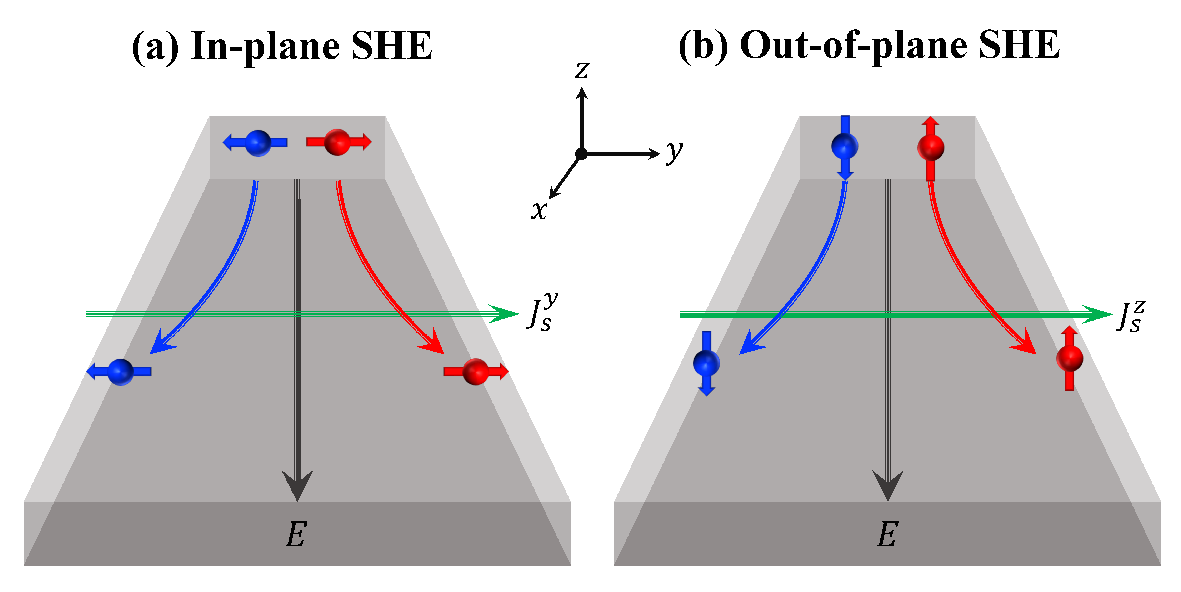}
\caption{ Schematic plots of the in-plane (a) and out-of-plane (b) spin Hall effects. $J_s^{y/z}$ denotes the spin current, and $E$ is the driving electric field. }\label{fig1}
\end{figure}

The results obtained from Bilbao Crystallographic Server, using Jahn Symbol $eV^3$, can be summarized as follows. For in-plane SHE (IP-SHE), we find: (1) for MPG m, m1', $\sigma^y_{xy} =\sigma^y_{yx} =0$ and $\sigma^x_{xy}$ and $\sigma^x_{yx}$ are indefinite; (2) for MPG 2', m', $\cal T$, both $\sigma^{x/y}_{xy}$ and $\sigma^{x/y}_{yx}$ are indefinite; (3) for MPG 3, 31', 3m', 6', $\sigma^{x/y}_{xy}=\sigma^{x/y}_{yx}$; (4) for MPG 3m, 3m1', 6'mm', $\sigma^{x}_{xy}=\sigma^{x}_{yx}=0$ and $\sigma^{y}_{xy}=\sigma^{y}_{yx}$. For other 2D MPG elements, both $\sigma^{x/y}_{xy}$ and $\sigma^{x/y}_{yx}$ are zero. For out-of-plane SHE (OP-SHE), we have: (1) $\sigma^{z}_{xy}=-\sigma^{z}_{yx}$ for most of the high symmetry rotations: 4, 41', 4mm, 4mm1', 4'm'm, 4m'm', 3, 31', 3m, 3m1', 3m', 6, 61', 6', 6mm, 6mm1', 6'mm', 6m'm'; (2) $\sigma^{z}_{xy}$ and $\sigma^{z}_{yx}$ are indefinite for the rest of the MPGs. The in-plane and out-of-plane SHEs are schematically shown in Fig.~\ref{fig1}.

These results are summarized in Table \ref{table2}, from which we find that for many systems the existence of SHE can not be determined solely by the symmetry. Moreover, we observe that $\sigma^{\alpha}_{xy}$ can be switched on and off while maintaining the symmetry of the system. For instance, $H_{\rm SOI} = k_x^2 \sigma_x + k_y^2 \sigma_y$ has $C_2 \cal T$ symmetry and it is easy to verify that $\sigma^{z}_{xy}$ is zero for $H_{\rm SOI}$ while $\sigma^{z}_{xy}\ne 0$ for $H_1 + H_{\rm SOI}$ where $H_1$ is the linear Rashba SOI listed in Table \ref{table1}. This shows that for a given Hamiltonian, symmetry alone can not characterize the OP-SHE. This conclusion is also valid for IP-SHE. Note that Eq.~(\ref{trans}) relates different components of SHE and imposes constraint on them (global constraint). It gives an upper bound of the number of independent components solely from the symmetry of the third rank pseudo-tensor, regardless of the physical system. Once the expression of SHE is given, we can use Eq.~(\ref{trans2}) to find further constraint on a particular component of SHE (local constraint). It turns out that the parity and types ($H_D$ or $H_R$) of SOI are good indicators to classify SHE. We recognize that parity is also a kind of symmetry, but in this work the word 'symmetry' refers solely to spatial symmetries. In the following, we will give the sufficient condition under which SHE may vanish.

\section{Classification of SHE based on the parity and types of SOIs}

The order of SOI, $N$, is determined by the power of $k$ (regardless of $k_x$ and $k_y$). In general, the parities of $H_D$ and $H_R$ are discussed in Appendix A. We find that the parity $(-1)^N$ plays a critical role in determining the existence of IP-SHE ($\sigma^{x/y}_{xy}$). Specifically, we demonstrate that, for a system with several SOI components, the IP-SHE can be switched on and off by tuning these SOIs while maintaining the symmetry of the system. We further show that SHE can be classified by the parity of SOI.


\renewcommand\arraystretch{1.2}
\begin{table}[tbp]
	\begin{center}
		\caption{\label{table2} Results from Neumann's principle for IP-SHE. Here "N" represents no constraint. "S" and "A" refer to symmetric relation for $\sigma^{x/y}_{xy}$ and antisymmetric relation for $\sigma^{z}_{xy}$ with respective to $x$ and $y$, respectively. "0" stands for zero $\sigma^\alpha_{xy}$. The results with $m_x$ are listed. For MPGs containing $m_y$, $\sigma^x_{xy}$ and $\sigma^y_{xy}$ are interchanged. For example, for MPG $m_y$, $m_y$1', $m_y$', $\sigma^x_{xy}$ = N and $\sigma^y_{xy}$ = 0.}
	\begin{tabular}{p{4cm}<{\centering}|p{1cm}<{\centering}|p{1cm}<{\centering}
|p{1cm}<{\centering}}
\hline\hline
MPG & $\sigma^x_{xy}$ & $\sigma^y_{xy}$ &$\sigma^z_{xy}$\\ \hline			
$m_x$, $m_x$1', $m_x$'& 0 & N & N\\ \hline
2'& N & 0 & N\\ \hline
$\cal T$ & N &N &N \\ \hline
3, 31', 6'&S &S &A   \\  \hline
3$m_x$, 3$m_x$1', 3$m_x$', 6'$m_y$'$m_x$ & 0 &S &A\\\hline
4mm, 4mm1', 4'm'm, 4m'm', 4, 41', 6mm1', 6mm, 6, 61', 6m'm'&0 &0 &A\\\hline
2, 21', 2mm, 2mm1', 2'm'm, 2m'm', 4' & 0 & 0 &N\\
\hline\hline			
		\end{tabular}
	\end{center}
\end{table}

\renewcommand\arraystretch{1.2}
\begin{table}[tbp]
	\begin{center}
		\caption{\label{table3} Results of direct calculation. Here "0" and "1" stand for zero and nonzero $\sigma^\alpha_{xy}$, respectively. "0/1" means SHE can be switched on and off and "NA" means that SOI Hamiltonian is not available for $d_z \ne 0$. The results with $m_x$ are listed. For MPGs containing $m_y$, $\sigma^x_{xy}$ and $\sigma^y_{xy}$ are interchanged.}
	\begin{tabular}{p{4cm}<{\centering}|p{1cm}<{\centering}|p{1cm}<{\centering}
|p{1cm}<{\centering}}
\hline\hline
MPG & $\sigma^x_{xy}$ & $\sigma^y_{xy}$ &$\sigma^z_{xy}$\\ \hline			
$m_x$& 0 & 1 & 0/1\\ \hline
$m_x$1', 3$m_x$1'& 0 & 1 & 1\\ \hline
$\cal T$, 3, 31' &0/1 &0/1 &1   \\  \hline
$m_x$', 3$m_x$,  3$m_x$' & 0 &0/1 &1\\\hline
4mm, 4'm'm, 4m'm', 4, 6mm, 6, 6m'm',
2, 2mm, 2'm'm, 2m'm', 4'   & 0 & 0 &1\\\hline
2', 21', 2'mm', 2mm1', 41', 4mm1', 6', 6'm'm, 61', 6mm1'& NA & NA& 1\\
\hline\hline			
		\end{tabular}
	\end{center}
\end{table}

For IP-SHE, there are 12 MPGs in Table \ref{table3}. The results are summarized as: (1) for MPG m, m1', 3m1, we find $\sigma_{xy}^x=0$ and $\sigma_{xy}^y=1$ ("1" stands for nonzero) when $m = m_x$; (2) for MPG m', 3m, 3m', we find $\sigma^{x}_{xy}=0$ while $\sigma^{y}_{xy}=0/1$ ("0/1" stands for zero or nonzero depending on the parity of individual SOI component in the Hamiltonian); (3) for MPG $\cal T$, 3, 31', $\sigma^{x/y}_{xy}=0/1$; (4) for MPG 2', 21', 2'mm', 2mm1', 41', 4mm1', 6', 6'm'm, 61', and 6mm1', no SOI Hamiltonians are available for 2D two-band models with nonzero $d_z$.

For OP-SHE, $\sigma^{z}_{xy}$ is nonzero for most of MPGs. The condition for $\sigma^{z}_{xy}=0$ is that $d_x d_y$ is an odd function in momentum space. We find that $\sigma^{z}_{xy}$ is nonzero for all 2D MPGs except three (m, 1, 2') with the following Hamiltonians: (1). the Hamiltonian with MPG m (mirror symmetry), e.g., $H = k_y^2 \sigma_x + k_x \sigma_y$. Other Hamiltonian with the same symmetry may not have vanishing OP-SHE. (2). the Hamiltonian with no symmetry at all, e.g., $H = k_x \sigma_x + k_y^2 \sigma_y$. (3). the Hamiltonian with MPG 2', e.g., $H = k_y^2 \sigma_x - k_x^2 \sigma_y$.
Another straightforward condition is that the system either has the chiral symmetry $\sigma_x$\cite{note11,note111} or chiral symmetry $\sigma_y$.\cite{note12} In the following, we give an example to demonstrate our findings and then verify it in a general account.


We consider the following Hamiltonian
\begin{eqnarray}
H &=& k^2\sigma_0 +\lambda_1(k_y\sigma_x-k_x\sigma_y)+ \dfrac{\lambda_2}{2} (k_+^{2}\sigma_+ + k_-^{2}\sigma_-) \nonumber \\
&-& i\dfrac{\lambda_3}{2} (k_+^{5}\sigma_+ - k_-^{5}\sigma_-)
+  \dfrac{\lambda_4}{2} (k^3_{+}+k^3_{-})\sigma_z
-\dfrac{i\lambda_5}{2} (k^3_{+} \nonumber \\
&-& k^3_{-})\sigma_z
+ \dfrac{\lambda_6}{2} (k^6_{+}+k^6_{-})\sigma_z - \dfrac{i\lambda_7}{2} (k^6_{+}- k^6_{-})\sigma_z. \label{tri}
\end{eqnarray}
Here $\sigma_{\pm} = \sigma_x \pm i\sigma_y$. In this Hamiltonian, different parts of SOI have been used before. For instance, the $\lambda_2$ term was used in Ref.~[\onlinecite{Acosta}] to address 2D dual topological insulator of Na$_3$Bi. By fitting the experimental data, the Dresselhaus-like SOI ($\lambda_3$ term) can account for the strong out-of-plane spin component at the Fermi surface of 2D Au/Ge(111) surface\cite{Hopfner}. The $\lambda_4$ term was proposed\cite{L-Fu3} to explain the experimentally observed warping effect of Fermi surface for topological insulator Bi$_2$Te$_3$ and the $\lambda_5$ term is crucial in determining the origin of experimental finding, the giant Zeeman-type spin polarization in WSe$_2$\cite{Yuan}. Note that the linear Rashba SOI ($\lambda_1$ term) has the highest symmetry while the $\lambda_2$ term (6'm'm), $\lambda_3$ term (6mm1'), $\lambda_6$ term (6), and $\lambda_7$ term (6mm) are hexagonal SOI. The $\lambda_4$ and $\lambda_5$ terms are trigonal SOI having $m_x$ and $m_y$ symmetries, respectively, making Eq.~(\ref{tri}) a good testing platform for symmetry analysis. For instance, when $\lambda_2$ and $\lambda_5$ are nonzero, the system has ($C_3$, $m_y \cal T$) symmetry (see Appendix B for more details). Turning on and off any of $\lambda_1$ \cite{suppress}, $\lambda_3$, or $\lambda_6$ does not affect the symmetry of the system. Fixing $\lambda_2$ and $\lambda_4$ to be nonzero while switching off $\lambda_5$ and $\lambda_6$ changes the system symmetry to ($C_3$, $m_x$). Therefore IP-SHE of this trigonal symmetry can be studied by tuning any of parameters of $\lambda_1$, $\lambda_3$, and $\lambda_7$. For nonzero $(\lambda_2$, $\lambda_4$, $\lambda_5)$, the system has $C_3$ symmetry. Its IP-SHE can be studied by varying $\lambda_1$. In addition, IP-SHE of hexagonal MPG 6 can be examined by setting $(\lambda_2$, $\lambda_4$, $\lambda_5)=0$ and$(\lambda_3$, $\lambda_6$, $\lambda_7) \neq 0$ while varying $\lambda_1$. IP-SHE of MPG 6mm (6m'm') can be studied by switching on $\lambda_3$ and $\lambda_7$ ($\lambda_6$) and turning on and off $\lambda_1$.

In the following, we focus on $c_\alpha \equiv \partial_x d_0 (\partial_y{\bf d}\times{\bf d})_\alpha$ which differs from $\Omega_{xy}^\alpha$ by $1/(4d^3)$. For convenience, we define $a_N = k_+^N + k_-^N$ and $b_N = i(k_+^N - k_-^N)$ so that all SOI Hamiltonians in Table \ref{table1} can be expressed in terms of $a_N$, $b_N$, and  $\sigma_\alpha$.
The parities of $a_N$ and $b_N$  with respect to $k_x$ and $k_y$ are found to be $a_N \sim k_x^N$ and $b_N \sim k_x k_y a_N$ (see Appendix C).
For Eq.~(\ref{tri}), in terms of $a_N$ and $b_N$, we have $d_x = \lambda_1 k_y + \lambda_2 a_2 + \lambda_3 b_5$, $d_y = -\lambda_1 k_x + \lambda_2 b_2 - \lambda_3 a_5$, and $d_z = \lambda_4 a_3 + \lambda_5 b_3 + \lambda_6 a_6 + \lambda_7 b_6$. Using Eq.~(\ref{spH}), the expression of $c_\alpha$ is shown in Appendix D.

Now we examine the behaviors of IP-SHE for various symmetries 6mm (6m'm'), 6, ($C_3$, $m_y \cal T$), ($C_3$, $m_x$) and $C_3$. (1). For MPG 6mm or 6m'm', we find $\sigma^{x/y}_{xy}=0$. The reason behind this can be understood by simple parity analysis. For a 2D system having two mirror symmetries or both $m_{x/y} \cal T$ symmetries, $d^2$ is an even function of both $k_x$ and $k_y$, and hence we can focus on the parity of $c_\alpha$. For $\sigma^\alpha_{xy}$ to be nonzero, $c_\alpha$ must be an even function of $k_0$ if we set $k_x=k_y=k_0$. In addition, the in-plane (out-of-plane) SOIs must be odd (even) functions of $k_0$. This is because $\sigma_{\pm}$ rotates in the same way as $k_{\pm}$ while $\sigma_z$ remains unchanged under rotation. Since $c_{x/y}$ scales like $d_{y/x} d_z k_x k_y$ according to Eq.~(\ref{spH}), it is impossible for $c_{x/y}$ to be an even function of $k_0$. Thus, we find that $\sigma^{x/y}_{xy}=0$ for systems with two mirror symmetries, which include symmetry groups: 2mm, 4mm, 6mm, 2mm1', 4mm1', 6mm1', 2m'm', 4m'm', and 6m'm'.

(2). For the system with $C_6$ symmetry, ($\lambda_3$, $\lambda_6$, $\lambda_7$) is nonzero and varying $\lambda_1$ does not affect the symmetry. We find $\sigma^{x/y}_{xy} =0$ because $d^2$ has a mirror symmetry $M_{x+y}$ with or without $\lambda_1$ so that terms involving $c^{(2)}_{x/y}$ and $c^{(3)}_{x/y}$ do not contribute to $\sigma^{x/y}_{xy}$. Discussion on the symmetry of energy spectrum or $d^2$ is shown in Appendix E. This result agrees with that obtained from Neumann's principle.

(3). For the system with ($C_3$, $m_y \cal T$) symmetry, we study the following cases. (a). The case when $\lambda_2$ and $\lambda_5$ are the only two nonzero parameters. Since there is only one SOI in either IP-SOI or OP-SOI, $d^2$ has two mirror symmetries. From Appendix D, we keep only $c^{(1)}_x$ and $c^{(1)}_y$ and find $\sigma^{x/y}_{xy}=0$ (no IP-SHE). As will be discussed later, if there is one IP-SOI and one OP-SOI with different parities, there is no IP-SHE. Case (3a) is just a special case. (b). Now we turn on $\lambda_1$ which respects $m_y$ symmetry. Since the $\lambda_2$ term does have $m_y$ symmetry, hence $d_x^2+d_y^2$ has only $m_x$ symmetry which is the only mirror symmetry possessed by both IP-SOIs. Note that the symmetry of $d_z^2$ remains the same as in (3a) which makes $d^2$ asymmetric about $k_y$. Including both $c^{(1)}_x$ and $c^{(3)}_x$, we obtain $c_x = 3\lambda_1\lambda_5 a_2 k_x^2 + \lambda_2\lambda_5 (2a_1 b_3-3a_2 b_2)k_x$ and $c_y=0$, which gives $\sigma^x_{xy} \neq 0$ and $\sigma^y_{xy} = 0$ (shown in Appendix F(a)). (c). If we replace the $\lambda_1$ term by the $\lambda_6$ term, $d_x^2+d_y^2$ has two mirror symmetries as explained above. Note that $d_z^2$ has $m_x$ symmetry. We have $c_x = -4 \lambda_2 \lambda_6 (a_1 a_6 + b_2 b_5)k_x + 2\lambda_2 \lambda_5 (3 a_2 b_2 - 2 a_1 b_3) k_x$ and $c_y = 0$, leading to nonvanishing $\sigma^{x}_{xy}$ (shown in Appendix F(b)). We see that $\sigma^x_{xy}$ and $\sigma^y_{xy}$ can be switched on and off while maintaining the system symmetry.

(4). For the system with ($C_3$, $m_x$) symmetry, we require $\lambda_2$ and $\lambda_4$ to be nonzero and set $\lambda_5= \lambda_6=0$. (a). When $\lambda_1= \lambda_7=0$, $d^2$ has two mirror symmetries and we find $\sigma^{x/y}_{xy}=0$ from the symmetry argument which is the same as case (3a). (b). When turning on $\lambda_1$, $d^2$ becomes symmetric about $k_x$ only. Hence neglecting terms odd in $k_x$ we find $\sigma^{x}_{xy}=0$ and $c_y = \lambda_1\lambda_4 (a_3 - 3b_2 k_y)k_x + \lambda_2\lambda_4 (2a_3b_1 - 3a_2 b_2)k_x$ making $\sigma^y_{xy}\ne 0$ (shown in Appendix F(c)). (c). When replacing the $\lambda_1$ term by the $\lambda_7$ term, the symmetry of $d^2$ remains the same. We obtain $\sigma^{x}_{xy}=0$ and $c_y=\lambda_2\lambda_4 (2a_3b_1 - 3a_2 b_2)k_x$ which has nonzero contribution. The result is the same as (3b). 

(5). When $(\lambda_2$, $\lambda_4$, $\lambda_5)$ are nonzero, the system has $C_3$ symmetry. It is easy to show that $d^2$ has only mirror symmetry $m_{x+y}$ and therefore only terms involving $c^{(1)}_{x/y}$ and $c^{(4)}_{x/y}$ contribute. We find $\sigma^{x/y}_{xy}=0$. When $\lambda_1$ is present, the system is still maintained at $C_3$ while $m_{x+y}$ symmetry is broken for $d^2$. We find $c_x = -6\lambda_1\lambda_5 a_2 k_x^2 + 6\lambda_1\lambda_4 b_2 k_x^2 - 2\lambda_2 \lambda_4 ( 2a_1a_3 + 3b_2^2)k_x + 2\lambda_2 \lambda_5 (3a_2 b_2 - 2 a_1 b_3)k_x$ and $c_y = 2\lambda_1\lambda_4 (a_3 - 3b_2 k_y)k_x + 2\lambda_1\lambda_5 (b_3 + 3a_2 k_y)k_x +2 \lambda_2 \lambda_5 (2b_1 b_3 + 3a_2^2)k_x + 2\lambda_2 \lambda_4 (2 a_3 b_1 - 3 a_2 b_2)k_x$. Hence both $\sigma^{x/y}_{xy} \ne 0$.

For OP-SHE, it is easy to see that: if OP-SHE vanishes for a particular $H_{SOI}$ then $H_1 + H_{SOI}$ makes $\sigma^z_{xy} \ne 0$. Therefore, for a given symmetry, OP-SHE can also be tuned from zero to nonzero. One IP-SOI means a basic building block with a defnite parity and SOI type, such as the one in Table \ref{table1}.


Now we give the sufficient conditions for vanishing IP-SHE in 2D systems in the following, and present the proof in Appendix I. (1). If the SOI Hamiltonian has chiral symmetry $\sigma_z$, then $\sigma^{x/y} _{xy} =0$. (2). If the SOI Hamiltonian contains only one in-plane and one out-of-plane components, there is no IP-SHE when the two componants have different parities (see the examples, cases (3a) and (4a)). Supposing they have the same parity, we find that $\sigma^{y}_{xy}=0$ if both components are the same type of SOI while $\sigma^{x}_{xy} = 0$ if they are different types of SOI. (3). The case that $d^2$ has one mirror symmetry $m_x$. Supposing the Hamiltonian is given by $H = H_0 + \lambda_1 R_{IP} + \lambda_2 D_{IP} + \lambda_3 D_{OP} + \lambda_4 R_{OP}$ where $R_{IP}$ and $D_{OP}$ stand for in-plane Rashba-like SOI and out-of-plane Dresselhaus-like SOI. Each SOI $A_B$ with $A = R,D$ and $B = IP,OP$ may have several terms but must have the same parity. For instance, $D_{IP} = D^a_{IP}+ D^b_{IP}$, IP-SHE vanishes only if $a$ and $b$ have the same parity. We find that the condition for $\sigma^{x}_{xy} = 0$ is: the parity of $(D_{IP}, R_{IP}, D_{OP}, R_{OP}) = (\pm, \mp, \mp, \pm)$ where $+$ stands for even parity while the condition for $\sigma^{y}_{xy} = 0$ is: the parity of $(D_{IP}, R_{IP}, D_{OP}, R_{OP}) = (\pm, \mp, \pm, \mp)$ (see the examples, cases (3b) and (4b-d)). (4). Assuming the mirror symmetry of $d^2$ is $m_y$, we find that the condition for $\sigma^x_{xy}=0$ is $H = R^-_{IP} + R^+_{IP}+ D^-_{OP} + D^+ _{OP}$ or $H = D^-_{IP} + D^+ _{IP}+ R^-_{OP} + R^+_{OP}$; while for $\sigma^y_{xy}=0$, we require $H = R^-_{IP} + R^+_{IP}+ R^-_{OP} + R^+_{OP}$ or $H = D^-_{IP} + D^+ _{IP}+ D^-_{OP} + D^+ _{OP}$. (5). If all IP-SOIs have one parity and all OP-SOIs have another parity, $\sigma^{x/y}_{xy}=0$.

We present examples of SOI Hamiltonians for all 2D MPGs in Table \ref{table5} of Appendix J. As shown in Appendix K, the vanishing of IP-SHE for all 2D MPGs with $d_z \ne 0$ in Table \ref{table3} can be predicted using conditions (2)-(5) without involving Neumann's principle.

\section{Discussion and conclusion}

In summary, we have demonstrated that the symmetry of the system along is not enough to characterize the existence of spin Hall effect, while the parity and symmetry types of constituent SOI play a crucial role. It is found that 2D in-plane SHE can be switched on and off by combining different SOI components, including in-plane and out-of-plane Rashba-like and Dresselhaus-like SOIs of different orders, while maintaining the system symmetry. Sufficient conditions for the existence of SHE are presented, accompanied by complete analysis on all 2D magnetic point groups.

The verification of our findings would be straightforward. SHE has been experimentally realized over fifteen years.\cite{Kato,Wunderlich} In Ref.~[\onlinecite{Basak}], $H = H_1 + H_9 +H_{13x}$ was used to model the surface states in the Bi$_2$Te$_3$ family of 3D TI using first-principles calculation, which has been experimentally verified.\cite{Hopfner}
However, we need a system with both types of SOI and different parities. In Ref.~[\onlinecite{Spielman}], a new class of atom-laser coupling schemes
were introduced to describe the spin-orbit coupled Hamiltonian of ultracold neutral atoms, which has the form $H = H_1 + H_{3x} +H_{3y} +H_5+H_6$ and could be adopted to the switching on and off of SHE.

\section*{Acknowledgement}
We acknowledge support from the National Natural Science Foundation of China (Grant Nos. 12034014, 12174262, and 12004442). L. Wang also thanks Guangdong Basic and Applied Basic Research Foundation (Grant No. 2021B1515130007) and Shenzhen Natural Science Fund (the Stable Support Plan Program 20220810130956001).

\section{Appendix}

\subsection{Symmetries of $H_D$ and $H_R$}

Under mirror symmetry $m_{x/y}$, we have $k_+ \rightarrow \mp k_-$ and $\sigma_+ \rightarrow \pm \sigma_-$, from which we find ${\tilde A} = (-1)^N A^\dagger$ under $m_x$ and ${\tilde A} = - A^\dagger$ under $m_y$ where ${\tilde A}$ is the corresponding SOI after transformation. Defining $(-1)^N$ as the parity of SOI, we obtain that $R^-_{IP}$ has both $m_{x/y}$, $R^+_{IP}$ has only $m_y$, $D^-_{IP}$ has no mirror symmetry while $D^+_{IP}$ has only $m_x$ symmetry, where $R$ and $D$ represent Rashba-like and Dresselhaus-like SOI, respectively. Since $\sigma_z \rightarrow -\sigma_z$ for $m_{x/y}$, we have ${\tilde B} = (-1)^{N+1} B^\dagger$ under $m_x$ and ${\tilde B} =  -B^\dagger$ under $m_y$, from which we find that $R^-_{OP}$ has $m_y$, $R^+_{OP}$ has both $m_{x/y}$, $D^-_{OP}$ has $m_x$, and $D^+_{OP}$ has no mirror symmetry (see Table \ref{table4} for a summary).

\renewcommand\arraystretch{1.2}
\begin{table}[tbp]
	\begin{center}
		\caption{\label{table4} Mirror symmetry of IP-SOI and OP-SOI. "A" and "S" represent antisymmetric and symmetric under different mirror operations, respectively. The SOI has the corresponding mirror symmetry if "S" is marked.}
	 \begin{tabular}{p{1.2cm}<{\centering}|p{0.8cm}<{\centering}|p{0.8cm}<{\centering}
|p{0.9cm}<{\centering}|p{1.2cm}<{\centering}|p{0.8cm}<{\centering}
|p{0.8cm}<{\centering}|p{0.9cm}<{\centering}}
\hline\hline
IP-SOI & $m_{x}$ & $m_{y}$ & $m_{x+y}$& OP-SOI & $m_{x}$& $m_{y}$& $m_{x+y}$\\ \hline			
$R^-_{IP}$ & S  & S & S& $R^-_{OP}$ &A  & S &A\\ \hline
$R^+_{IP}$& A & S & A&$R^+_{OP}$ &S & S&S\\ \hline
$D^-_{IP}$& A & A & S& $D^-_{OP}$ & S& A&A\\ \hline
$D^+_{IP}$& S & A & A& $D^+_{OP}$ & A& A&S\\
\hline\hline			
		\end{tabular}
	\end{center}
\end{table}

\subsection{Symmetry of the Hamiltonian in Eq.~(\ref{tri})}

For Eq.~(\ref{tri}), the symmetry of each SOI term is listed:
$\lambda_1$ term has ($C_\infty$, $\cal T$);
$\lambda_2$ term has ($C_6 \cal T$, $C_3$, $m_x$, $m_y \cal T$);
$\lambda_3$ term has ($C_6$, $\cal T$, $C_3$, $m_x$, $m_y$, $m_x \cal T$, $m_y \cal T$);
$\lambda_4$ term has ($C_3$, $\cal T$, $m_x$, $m_x \cal T$);
$\lambda_5$ term has ($C_3$, $\cal T$, $m_y$, $m_y \cal T$);
$\lambda_6$ term has ($C_6, C_3, m_x\mathcal{T},m_y\mathcal{T}$);
$\lambda_7$ term has ($C_6$, $C_3$, $Mmx$, $m_y$).

We may have three cases to test the effect of symmetry on IP-SHE. (1). The presence of $\lambda_3$ and $\lambda_6$ ensures $C_6$ symmetry and the presence of $\lambda_3$ and $\lambda_7$ ensures ($C_6$, $m_x$, $m_y$) symmetries. (2). The presence of $\lambda_2$ and $\lambda_5$ ensures ($C_3$, $m_y \cal T$) symmetries regardless of $\lambda_1$, $\lambda_3$, and $\lambda_4$. (3). The presence of $\lambda_2$ and $\lambda_4$ while setting $\lambda_5=\lambda_6=0 $ ensures $C_3$ and $m_x$ symmetries. Note that varying ($\lambda_1$, $\lambda_3$, $\lambda_7$) does not change the symmetry of the system. All of these systems do not preserve time-reversal symmetry.

\subsection{Parity of $a_N$ and $b_N$}


We define $a_N = k_+^N + k_-^N$ and $b_N = i(k_+^N - k_-^N)$.
The parities of $a_N$ and $b_N$ can be obtained as follows. Under $m_x$, we have $k_+ \rightarrow -k_-$, $a_N \rightarrow (-1)^{N}a_N$ and $b_N \rightarrow (-1)^{N+1}b_N$. Under $m_y$, we have $k_+ \rightarrow k_-$, $a_N \rightarrow a_N$ and $b_N \rightarrow -b_N$. Therefore $a_N \sim k_x^N g_1$ and $b_N \sim k_x^{N-1} k_y g_2$ where $g_i$ with $i=1,2$ are arbitrary even functions of $k_\alpha$.


\subsection{Analytic expressions of $c_\alpha$}

$c_\alpha$ is given by
\begin{eqnarray}
c_z &=& \partial_x d_0 (d_x \partial_y d_y - d_y \partial_y d_x), \nonumber \\
c_{x/y} &=& \partial_x d_0 (d_z \partial_y d_{y/x} - d_{y/x} \partial_y d_z).  \label{cz}
\end{eqnarray}
For the example in the main text, $c_z$ is found to be
\begin{eqnarray}
c^{(1)}_z &=& 2 [\lambda_1^2 k_x + \lambda_1 \lambda_3 (a_5 -5a_4 k_x-5b_4 k_y)\nonumber \\
&-& 2\lambda_2^2 (b_1 b_2 + a_1 a_2) -5\lambda_3^2 (a_4 a_5 + b_4 b_5)]k_x, \nonumber \\
c^{(2)}_z &=&0, \nonumber \\
c^{(3)}_z &=& 2[\lambda_1 \lambda_2 (2 b_1 k_x-2a_1 k_y-b_2) + \lambda_2 \lambda_3 (2 a_5 b_1 \nonumber \\
&-&5a_2 b_4 -2a_1 b_5 + 5a_4 b_2)]k_x, \nonumber \\
c^{(4)}_z &=&0,
\end{eqnarray}
where $c^{(1)}_z$ is even in $k_x$ and $k_y$, $c^{(2)}_z$ is odd in $k_x$ and even in $k_y$, $c^{(3)}_z$ is even in $k_x$ and odd in $k_y$, and $c^{(4)}_z$ is odd in $k_x$ and $k_y$. We have used the relations $\partial_y a_n = n b_{n-1}$ and $\partial_y b_n = - n a_{n-1}$.
$c_x$ also has four terms,
\begin{eqnarray}
c^{(1)}_x &=& 2 [-3 \lambda_1 \lambda_5 a_2 k_x - \lambda_2 \lambda_6 (2 a_1 a_6 + 6 b_2 b_5) \nonumber \\
&-& \lambda_3 \lambda_5 (5b_3 b_4 +3 a_2 a_5)]k_x, \nonumber \\
c^{(2)}_x &=& 2[ \lambda_2 \lambda_4 (- 2a_1a_3 - 3b_2^2)
-6 \lambda_1 \lambda_7 a_5 k_x \nonumber \\
&-&\lambda_3\lambda_7(6a_5^2+5b_4b_6) ] k_x, \nonumber \\
c^{(3)}_x &=& 2 [\lambda_2 \lambda_5 (3a_2 b_2 - 2 a_1 b_3) - \lambda_3 \lambda_6 (5 a_6 b_4 \nonumber \\
&-&6 a_5 b_5)+ 6\lambda_1 \lambda_6 b_5 k_x]k_x, \nonumber \\
c^{(4)}_x &=& 2 [ \lambda_3 \lambda_4(3 a_5 b_2 - 5a_3 b_4) +3\lambda_1 \lambda_4 b_2 k_x \nonumber \\
&+& \lambda_2\lambda_7(6a_5b_2-2a_1b_6)]k_x. \label{cx}
\end{eqnarray}
For $c_y$, we find
\begin{eqnarray}
c^{(1)}_y &=& 2[\lambda_1 \lambda_4 (a_3 -3 b_2 k_y) - \lambda_3 \lambda_4 (5a_3 a_4 + 3b_2 b_5) \nonumber \\
&+&\lambda_2\lambda_7(6a_2a_5+2b_1b_6)]k_x, \nonumber \\
c^{(2)}_y &=& 2 [\lambda_1 \lambda_6 (a_6 -6 b_5 k_y) + \lambda_2 \lambda_5 (2b_1 b_3 + 3a_2^2) \nonumber \\
&+& \lambda_3 \lambda_6 (5 a_4 a_6 -6b_5^2)]k_x, \nonumber \\
c^{(3)}_y &=& 2 [\lambda_2 \lambda_4 (2 a_3 b_1 - 3 a_2 b_2) + \lambda_1\lambda_7(b_6+6a_5k_y) \nonumber \\
&+& \lambda_3\lambda_7 (6a_5 b_5 - 5 a_4 b_6) ]k_x,\nonumber \\
c^{(4)}_y &=& 2 [\lambda_1 \lambda_5 (b_3 + 3 a_2 k_y) + \lambda_2 \lambda_6 (2 a_6 b_1 -6 a_2 b_5)\nonumber \\
&-& \lambda_3 \lambda_5 (5 a_4 b_3 - 3 a_2 b_5)]k_x. \label{cy}
\end{eqnarray}


\subsection{Symmetry of the energy spectrum}


For $H_0 = k^2 \sigma_0$, $d^2$ is the same as that of the energy spectrum, from the symmetry point of view. The symmetry of the energy spectrum is summarized as follows.

(a). $m_x$ and $m_y$ symmetries. If there is only one IP-SOI and one OP-SOI, the energy spectrum has two mirror symmetries regardless of symmetries of each SOI.

(b). $m_{x+y}$ symmetry. If all IP-SOIs have one parity and all OP-SOIs have another parity, the energy spectrum has $m_{x+y}$ symmetry.

(c). $m_x$ symmetry. If IP-SOI $H_{IP} = R^-_{IP} + D^+_{IP}$ or $R^+_{IP} + D^- _{IP}$ then $d_x^2 +d_y^2$ has $m_x$ symmetry, where $R^-_{IP}$ denotes Rashba-type of IP-SOI with odd parity. If OP-SOI $H_{OP} = R^-_{OP} + D^+_{OP}$ or $R^+_{OP} + D^- _{OP}$ then $d_z^2$ has $m_x$ symmetry. Hence there are four possible combinations that respect $m_x$ symmetry. Note that $H_{IP} = R^-_{IP} + R^-_{IP} + D^+_{IP} = R^-_{IP} + D^+_{IP}$.

(d). $m_y$ symmetry. If IP-SOI $H_{IP} = R^-_{IP} + R^+_{IP}$ or $D^-_{IP} + D^+ _{IP}$ then $d_x^2 +d_y^2$ has $m_y$ symmetry. If OP-SOI $H_{OP} = R^-_{OP} + R^+_{OP}$ or $D^-_{OP} + D^+ _{OP}$ then $d_z^2$ has $m_y$ symmetry. Once again there are four possible combinations that respect $m_y$ symmetry.


\subsection{Details of calculation for the examples in the main text}


(a). For the system with ($C_3$, $m_y \cal T$) symmetry, when ($\lambda_1$, $\lambda_2$, $\lambda_5$) are nonzero, $\int_k c_x/d^3 \ne 0$. It is straightforward to find $d^2 = \lambda_1^2 k^2 + \lambda_2^2 k^4 + \lambda_1 \lambda_2 k^3 \sin (3\theta) + \lambda_5^2 \sin^2(3\theta)$ and $c_x = 3\lambda_1\lambda_5 k^4 \cos2\theta \cos^2\theta + \lambda_2\lambda_5 k^6 (2\cos\theta\sin3\theta-3\cos2\theta \sin2\theta)\cos\theta = (3/4) \lambda_1\lambda_5 k^4 (1+2\cos2\theta +\cos4\theta) + (1/4) \lambda_2\lambda_5 k^6 (7\sin3\theta +5\sin5\theta -2\sin\theta)$ where we have used $a_N = k^N \cos N\theta$ and $b_N = -k^N \sin N\theta$. Expanding $d^2$ in powers of $\sin3\theta$, the $\sin^2(3\theta)$ term gives a nonzero contribution in $\lambda_1\lambda_5$ while the $\sin3\theta$ term makes $\lambda_2\lambda_5$ nonzero.

(b). For the system with ($C_3$, $m_y \cal T$) symmetry, when ($\lambda_2$, $\lambda_5$, $\lambda_6$) are nonzero, $\int_k c_x/d^3 \ne 0$. Note that $c_x = -4 \lambda_2 \lambda_6 k^8 (\cos\theta \cos6\theta + \sin2\theta \sin5\theta)\cos\theta + 2 \lambda_2 \lambda_5 k^5 (3/2 \sin4\theta -2\cos\theta \sin3\theta) \cos\theta =- \lambda_2 \lambda_6 k^8 (2\cos8\theta +5 \cos6\theta -\cos2\theta)- (1/2) \lambda_2 \lambda_5 k^5 (\sin5\theta + \sin3\theta + 4\sin\theta)$. Since $d^2$ is the same as (a), the $\cos6\theta$ and $\sin3\theta$ terms give nonzero contributions to $\sigma^x_{xy}$.

(c). For the system with ($C_3$, $m_x$) symmetry, when ($\lambda_1$, $\lambda_2$, $\lambda_4$) are nonzero, $\int_k c_y/d^3 \ne 0$. Note that $d^2 = \lambda_1^2 k^2 + \lambda_2^2 k^4 + \lambda_1 \lambda_2 k^3 \sin (3\theta) + \lambda_4^2 \cos^2(3\theta)$ and $c_y = \lambda_1\lambda_4 k^4 (\cos3\theta + 3\sin2\theta \sin\theta)\cos\theta + \lambda_2\lambda_4 k^5(-2\cos3\theta \sin\theta + 3\cos2\theta \sin2\theta)\cos\theta = (1/4) \lambda_1\lambda_4 k^4 (2\cos4\theta -\cos2\theta +3) + (1/4) \lambda_2\lambda_4 k^5 (\sin5\theta +3\sin3\theta +2 \sin\theta)$ which is nonzero due to the constant term in $\lambda_1\lambda_4$ and $\sin3\theta$ term in $\lambda_2\lambda_4$.

\subsection{Parity analysis for SHE}


If we are only interested in the parity of $c_\alpha$, parity analysis is most convenient. We have
\begin{eqnarray}
c_z &=& k_x k_y d_x d_y, \nonumber \\
c_x &=& k_x k_y d_y d_z, \nonumber \\
c_y &=& k_x k_y d_x d_z, \label{parity}
\end{eqnarray}
where the factor $k_y$ comes from $\partial_y$. Now we take the example in the main text to demonstrate this method. Let ($\lambda_1$, $\lambda_2$, $\lambda_5$) be nonzero, we find
\begin{eqnarray}
c_x &=& k_x k_y(-\lambda_1 \lambda_5 k_x b_3 + \lambda_2 \lambda_5 b_2 b_3) \nonumber \\
&=& \lambda_1 \lambda_5 g_1 + \lambda_2 \lambda_5 k_y g_2, \nonumber
\end{eqnarray}
and
\begin{eqnarray}
c_y &=& k_x k_y(\lambda_1 \lambda_5 k_y b_3 + \lambda_2 \lambda_5 a_2 b_3) \nonumber \\
&=& \lambda_1 \lambda_5 k_x k_y g_1 + \lambda_2 \lambda_5 k_x g_2, \nonumber
\end{eqnarray}
where $g_1$ and $g_2$ are even functions of $k_\alpha$. The same results are found from Eqs.~(\ref{cx}) and (\ref{cy}) if we concern only about the parity. Note that this parity analysis is useful in proving sufficient conditions for zero IP-SHE.

\subsection{Existence of OP-SHE}


We show that OP-SHE can be nonzero for all 2D MPGs. Since the linear Rashba SOI $H_1$ has ($C_\infty$, $m_x$, $m_y$, $\cal T$) symmetry, for any SOI Hamiltonian with a particular MPG $H_{\rm MPG} = d_1\sigma_x+ d_2\sigma_y +d_3\sigma_z$, obviously $H_{\rm MPG} + H_1$ will not change the symmetry of the system. Therefore, we can write $d_x = \lambda_1 k_y + \lambda_2 d_1$, $d_y = -\lambda_1 k_x + \lambda_2 d_2$, $d_z = \lambda_2 d_3$. According to the expression of $\Omega_{xy}^z$ in Eq.~(\ref{BC}), we have
\begin{eqnarray}
\Omega_{xy}^z &=& [\lambda_1^2 k_x^2 + \lambda_2^2k_x (d_1 \partial_y d_2 - d_1 \partial_y d_2) \nonumber \\
&+& \lambda_1\lambda_2 k_x (k_y \partial_y d_2 + k_x \partial_y d_1 - d_2)]/(4 d^3 ).
\end{eqnarray}
It is obvious that the $\lambda_1^2$ term already contributes to a nonzero value of $\sigma^z_{xy}$, and other terms will not make it vanish for general $\lambda_1$ and $\lambda_2$.

\subsection{Proof of the conditions for zero IP-SHE}

%

\noindent{\it Condition (1)} --- If SOI contains no out-of-plane component, it has chiral symmetry $\sigma_z$. Consequently, $\sigma^{x/y} _{xy} =0$ according to Eq.~(\ref{cz}).

\medskip

\noindent{\it Condition (2)} --- If SOI contains only one in-plane and one out-of-plane components (with $H_0 = k^2\sigma_0$), $d^2$ has two mirror symmetries. So we keep only terms that are even in $k_x$ and $k_y$. If IP-SOI and OP-SOI have different parities, $d_{x/y} d_z$ must have odd power of $k_0 = k_x=k_y$. Therefore $c_{x/y} \sim k_x k_y d_{y/x} d_z$ is an odd function of $k_0$ making $\sigma^{x/y}_{xy} \sim \int_k c_{x/y}/d^3 = 0$.

If IP-SOI and OP-SOI have the same parity, we consider the following four combinations.

(a). For $D_{IP}$ and $D_{OP}$ (one from in-plane SOI and the other from out-of-plane), $c_{x/y}$ can be analyzed from their parities. From Table \ref{table4} in Appendix A, we see that both $D_{IP}$ and $D_{OP}$ are antisymmetric under $m_y$ regardless of the parity. Since under $m_y$, $k_y \rightarrow -k_y$ and $\sigma_y \rightarrow \sigma_y$ while $\sigma_{x/z} \rightarrow -\sigma_{x/z}$, we obtain that $d_x$ ($d_y$) is an even (odd) function of $k_y$ and $d_z$ is an even function of $k_y$. Hence $c_y \sim k_x k_y d_x d_z$ is odd in both $k_x$ and $k_y$ leading to $\sigma^{y}_{xy}=0$.


(b). For $R_{IP}$ and $R_{OP}$, Table \ref{table4} shows that both $R_{IP}$ and $R_{OP}$ are symmetric under $m_y$ regardless of the parity. Similar to the argument of (a), we find that $d_x$ ($d_y$) is an odd (even) function of $k_y$ and $d_z$ is an odd function of $k_y$. Once again $c_y \sim k_x k_y d_x d_z$ is odd in both $k_x$ and $k_y$ making $\sigma^{y} _{xy}=0$.


(c). For $D_{IP}$ and $R_{OP}$, Table \ref{table4} shows that $D_{IP}$ ($R_{OP}$) is antisymmetric (symmetric) under $m_y$ regardless of the parity. Similar to the argument of (a), we find that $d_x$ ($d_y$) is an even (odd) function of $k_y$ and $d_z$ is an odd function of $k_y$. Thus $c_x \sim k_x k_y d_y d_z$ is odd in both $k_x$ and $k_y$ making $\sigma^{x} _{xy}=0$.


(d). For $R_{IP}$ and $D_{OP}$, Table \ref{table4} shows that $R_{IP}$ ($D_{OP}$) is symmetric (antisymmetric) under $m_y$ regardless of the parity. Similar to the argument of (a), we find that $d_x$ ($d_y$) is an odd (even) function of $k_y$ and $d_z$ is an even function of $k_y$. Thus $c_x \sim k_x k_y d_y d_z$ is odd in both $k_x$ and $k_y$ making $\sigma^{x}_{xy}=0$.


\renewcommand\arraystretch{1.2}
\begin{table*}[tbp]
	\begin{center}
		\caption{\label{table5} SOI Hamiltonians and corresponding symmetries for 31 MPGs.}
		\begin{tabular}{p{4.1cm}<{\centering}|p{4.1cm}<{\centering}|
p{5cm}<{\centering}|p{4.2cm}<{\centering}}
\hline\hline
$H_1 = {\Im} (k_- \sigma_+)$ ($C_\infty$, $\cal T$) & 			
$H_2= {\Re} (k_+\sigma_+)$  (21')  & $H_{3x}= {\Re}(k_+^2\sigma_-)$  (2'mm')& $H_{3y} =  {\Im}(k_+^2\sigma_-)$ (2'm'm)\\ \hline
$H_{4x} = {\Re}(k_+^2\sigma_+)$  (6'mm')& $H_{4y} =  {\Im}(k_+^2\sigma_+)$ (6'm'm) & $H_5 = {\Re}(k_+^3\sigma_+)$  (41')&$H_6 =  {\Im}(k_+^3\sigma_-)$  (2mm1')\\ \hline
$H_{7} =  {\Im}(k_+^3\sigma_+)$  (4mm1')& $H_{8} = {\Re}(k_+^5\sigma_+)$ (61')&$H_{9} =  {\Im}(k_+^5\sigma_+)$  (6mm1')& $H_{10x} = {\Re}(k_+) \sigma_z$  (m1') \\ \hline
$H_{10y} =  {\Im}(k_+) \sigma_z$ (m1')& $H_{11} = {\Re}(k_+^2) \sigma_z$ (4'm'm) & $H_{12} =  {\Im}(k_+^2) \sigma_z$  (4'm'm)&$H_{13x} = {\Re}(k_+^3) \sigma_z$ (3m1')\\ \hline
$H_{13y} =  {\Im}(k_+^3) \sigma_z$  (3m1')& $H_{14} = {\Re}(k_+^4) \sigma_z$ (4m'm') &$H_{15} =  {\Im}(k_+^4) \sigma_z$  (4mm)& $H_{16} = {\Re}(k_+^6) \sigma_z$ (6m'm')\\ \hline
$H_{17} =  {\Im}(k_+^6) \sigma_z$  (6mm) & $H_{18} =H_2 +H_3$ (2') &$H_{19} =H_1 + H_{4x} +H_{4y}$ (6')&$H_{20x} =H_1 + H_{3y} +H_{11}$ (m')\\ \hline
$H_{20y} =H_1 + H_{3x} +H_{11}$ (m')& $H_{21} =H_2 + H_{12}$ (2)&$H_{22} =H_1 + H_{11}+ H_{14}$ (2m'm')&$H_{23} =H_5 + H_{14}$ (4)\\ \hline
$H_{24x} =H_1 +H_{4y}+ H_{16} $ (3m')&$H_{24y} =H_1 +H_{4x} + H_{16} $ (3m')&$H_{25} =H_{4x} +H_{4y} + H_{13x} $ (3)&$H_{26} =H_1 + H_{16} +H_{17} $ (6)\\ \hline
$H_{27} =H_{1} +H_{10x} + H_{10y} $ ($\cal T$)&$H_{28x} =H_{1} +H_{3x} + H_{12} $ (m)&$H_{28y} =H_{1} +H_{3y} + H_{12} $ (m)&$H_{29} =H_{1} +H_{12} + H_{15} $ (2mm)\\ \hline
$H_{30} =H_{1} +H_{11} + H_{12} $ (4')&$H_{31} =H_{1} +H_{13x} + H_{13y} $ (31')& $H_{32x} =H_{1} +H_{4x} + H_{13x} $ (3m)& $H_{32y} =H_{1} +H_{4y} + H_{13y} $ (3m)\\
\hline\hline	
$H'_2 = {\Re}(k_+^3\sigma_-)$ (21')& $H'_{4x} = {\Re}(k_+^4\sigma_-)$ (6'mm')&$H'_{4y} =  {\Im}(k_+^4\sigma_-)$  (6'm'm)&$H'_{5} = {\Re}(k_+^5\sigma_-)$ (41') \\ \hline
$H'_{7} =  {\Im}(k_+^5\sigma_-)$  (4mm1')&\\
\hline\hline		
		\end{tabular}
	\end{center}
\end{table*}

\medskip
\noindent{\it Condition (3)} --- Since $d^2$ has $m_x$ symmetry, from section I (2c) we anticipate the combination of SOIs to be $H = k^2\sigma_0 + \lambda_1 D^l_{IP} + \lambda_2  R^k_{IP}+ \lambda_3 D^m_{OP} + \lambda_4 R^n_{OP}$ with $d_x = \lambda_1 a_l +\lambda_2 b_k$, $d_y = \lambda_1 b_l +\lambda_2 a_k$, and $d_z = \lambda_3 a_m +\lambda_4 b_n$. From Eq.(\ref{parity}), we have
\begin{eqnarray}
c_x &\sim& k_xk_y(\lambda_1 \lambda_3 k_y k_x^{l+m+1} +\lambda_1 \lambda_4 k_x^{l+n} \nonumber \\
&+& \lambda_2 \lambda_3 k_x^{k+m} +\lambda_2 \lambda_4 k_y k_x^{k+n+1}), \label{q1}
\end{eqnarray}
and
\begin{eqnarray}
c_y &\sim& k_x k_y(\lambda_1 \lambda_3 k_x^{l+m} +\lambda_1 \lambda_4 k_y k_x^{l+n+1} \nonumber \\
&+& \lambda_2 \lambda_3 k_y k_x^{k+m+1} +\lambda_2 \lambda_4 k_x^{k+n}). \label{q2}
\end{eqnarray}

Since $d^2$ has $m_x$ symmetry, we can ignore those terms odd in $k_x$. Hence the sufficient conditions for $\sigma^{x}_{xy} = 0$ are $l+m=$ odd, $l+n=$ even, $k+m=$ even, and $k+n=$ odd. In another word, we have two solutions:
\begin{eqnarray}
{\rm  parity ~of} ~(D_{IP}, R_{IP}, D_{OP}, R_{OP}) = (\pm, \mp, \mp, \pm), \label{sigmax}
\end{eqnarray}
where $+$ labels even parity. The solutions for $\sigma^{y}_{xy} = 0$ are:
\begin{eqnarray}
{\rm  parity ~of} ~(D_{IP}, R_{IP}, D_{OP}, R_{OP}) = (\pm, \mp, \pm, \mp). \label{sigmay}
\end{eqnarray}
Note that if we only have three SOI components, e.g., $R_{IP}=0$, Eqs.~(\ref{sigmax}) and (\ref{sigmay}) can still be used. In this case, Eq.(\ref{sigmax}) reads: ${\rm parity ~of} ~(D_{IP},  D_{OP}, R_{OP}) = (\pm, \mp, \pm)$. The combination of SOIs in Eqs.~(\ref{sigmax}) and (\ref{sigmay}) can also be obtained from Table \ref{table4} from the symmetry point of view. We illustrate this method in the analysis of condition (4) below.

\medskip
\noindent{\it Condition (4)} --- Since $d^2$ has $m_y$ symmetry, from section I (2c) we focus on the following four combinations of SOI.

(a). $H_{IP} = R^-_{IP} + R^+_{IP}$ and $H_{OP} = R^-_{OP} + R^+_{OP}$. From Table \ref{table4}, we know that both $H_{IP}$ and $H_{OP}$ are symmetric under $m_y$. Therefore, $d_x$ ($d_y$) is antisymmetric (symmetric) about $k_y$ and $d_z$ is antisymmetric about $k_y$ because under $m_y$ only $\sigma_x$ changes sign while $\sigma_y$ does not. As a result, $c_y \sim k_x k_y d_x d_z$ is an odd function of $k_y$ giving rise to $\sigma^y_{xy}=0$.

(b). $H_{IP} = R^-_{IP} + R^+_{IP}$ and $H_{OP} = D^-_{OP} + D^+ _{OP}$. From Table \ref{table4}, we know that $H_{IP}$ ($H_{OP}$) is symmetric (antisymmetric) under $m_y$. Therefore, $d_x$ ($d_y$) is antisymmetric (symmetric) about $k_y$ and $d_z$ is symmetric about $k_y$. As a result, $c_x \sim k_x k_y d_y d_z$ is an odd function of $k_y$ making $\sigma^x_{xy}=0$.

(c). $H_{IP} = D^-_{IP} + D^+ _{IP}$ and $H_{OP} = R^-_{OP} + R^+_{OP}$. From Table \ref{table4}, we know that $H_{IP}$ ($H_{OP}$) is antisymmetric (symmetric) under $m_y$. Therefore, $d_x$ ($d_y$) is symmetric (antisymmetric) about $k_y$ and $d_z$ is antisymmetric about $k_y$. As a result, $c_x \sim k_x k_y d_y d_z$ is an odd function of $k_y$ making $\sigma^x_{xy}=0$.

(d). $H_{IP} = D^-_{IP} + D^+ _{IP}$ and $H_{OP} = D^-_{OP} + D^+ _{OP}$. From Table \ref{table4}, we know that both $H_{IP}$ and $H_{OP}$ are antisymmetric under $m_y$. Therefore, $d_x$ ($d_y$) is symmetric (antisymmetric) about $k_y$ and $d_z$ is symmetric about $k_y$. As a result, $c_y \sim k_x k_y d_x d_z$ is an odd function of $k_y$ making $\sigma^y_{xy}=0$.

Thus, the condition for $\sigma^x_{xy}=0$ is $H = R^-_{IP} + R^+_{IP}+ D^-_{OP} + D^+ _{OP}$ or $H = D^-_{IP} + D^+ _{IP}+ R^-_{OP} + R^+_{OP}$ while for $\sigma^y_{xy}=0$, we require $H = R^-_{IP} + R^+_{IP}+ R^-_{OP} + R^+_{OP}$ or $H = D^-_{IP} + D^+ _{IP}+ D^-_{OP} + D^+ _{OP}$.

\medskip
\noindent{\it Condition (5)} --- For the case where $d^2$ has $m_{x+y}$ symmetry. If all IP-SOIs have one parity and all OP-SOIs have another parity, then $d^2$ has $m_{x+y}$ symmetry. We see from Table \ref{table4} that, if the parity of IP-SOI is positive (negative), $d_{x/y}$ are symmetric (antisymmetric) under $m_{x+y}$ because $\sigma_{x/y}$ also change sign. For OP-SHE, it is the opposite, i.e., if parity of OP-SOI is positive (negative), $d_{x/y}$ are antisymmetric (symmetric) under $m_{x+y}$. As a result, if the parities of IP-SOI and OP-SOI are different, $d_{y/x} d_z$ is an even function of $k_y$ and $c_{x/y} \sim k_x k_y d_{y/x} d_z$ is an odd function. Thus $\sigma^{x/y}_{xy}=0$.


\subsection{List of SOI Hamiltonians for 31 MPGs}

Here we list SOI Hamiltonians for 31 MPGs in Table \ref{table5}. Note that the construction of a SOI Hamiltonian for a given symmetry is not unique. They can have different expressions with the same symmetry. For instance, $H_1 + H_2$ has the same symmetry as that of $H_2$. $H_{11}$ has $m_x \cal T$, $m_y \cal T$ as well as  $m_{x+y}$. On the other hand, $H_{12}$ has $m_x$, $m_y$, and $m_{x+y}$. Here under $m_{x+y}$, $k_{x/y} \rightarrow -k_{x/y}$ and $\sigma_z$ remains unchanged. Within our 2D two-band model, only 20 MPGs have nonzero $d_z$ and hence can be used to study IP-SHE.

\subsection{Analysis of IP-SHE for all 2D MPGs using the sufficient conditions}

In the following, we examine the IP-SHE for all 2D MPGs with $d_z \ne 0$ that are listed in Table III of the main text.

\medskip
\noindent{\bf MPG m, m1', 3m1'} --- For these three MPGs with m = $m_x$, $\sigma^{x} _{xy} = 0$ and $\sigma^{y}_{xy} = 1$ according to the notation of Table III. Now we analyze these MPGs in details.

(a). MPG m. For $H_{28x} = R^-_{IP} + D^+_{IP} + R^+_{OP}$ with $m_x$, we have $\sigma^{x}_{xy} = 0$ from condition (3). From direct calculation, we find that $c_y$ contains $\sin3\theta$ and $d^2 = A + B \sin(3\theta) + C \cos(4\theta)$. Thus $\sigma^{y}_{xy} \ne 0$. For $H_{28y} = R^-_{IP} + R^+_{IP} + R^+_{OP}$ with $m_y$, we find $c_y$ is exactly zero. Direct calculation of $c_x$ is the same as $c_y$ of $H_{28x}$ and thus $\sigma^{x}_{xy} \ne 0$ and $\sigma^{y}_{xy} = 0$.

(b). MPG m1'. For $H_{10x} = R^-_{IP} + D^-_{OP}$ with $m_x$1', we have $\sigma^x_{xy}=0$ from condition (2). Direct calculation gives $c_y = k_x^2$ and thus $\sigma^y_{xy} \ne 0$. For $H_{10y} = R^-_{IP} + R^-_{OP}$ with $m_y$1', we have $\sigma^y_{xy}=0$ from condition (2). Direct calculation gives $c_x = k_x^2$ and thus $\sigma^x_{xy} \ne 0$.

(c). MPG 3m1'. For $H_1+ H_{13x} = R^-_{IP} + D^-_{OP}$ with 3$m_x$1', we have $\sigma^x_{xy}=0$ from condition (2). Direct calculation gives $c_y = k_x^4$ and thus $\sigma^y_{xy} \ne 0$. For $H_1+ H_{13y} = R^-_{IP} + R^-_{OP}$ with 3$m_y$1', we have $\sigma^y_{xy}=0$ from condition (2). Direct calculation gives $c_x = 3k^2_x (k_y^2 -k_x^2) \sim 1+ \cos2\theta + \cos4\theta$ and thus $\sigma^x_{xy} \ne 0$ due to the constant term.

\medskip
\noindent{\bf MPG $\cal T$, 3, 31'} --- For these three MPGs, $\sigma^{x/y}_{xy} = 0/1$ according to the notation of Table III. Now we analyze these MPGs in details.

(a). MPG $\cal T$. For $H_{27} = R^-_{IP} + D^-_{OP} + R^-_{OP}$ with $\cal T$, direct calculation gives $c_x = c_y = k_x^2$ and hence $\sigma^{x/y}_{xy} \ne 0$. For $H_A = H_2 + H_{13x} = D^-_{IP} + D^-_{OP}$ and $H_A = H_2 + H_{13y} = D^-_{IP} + R^-_{OP}$, the system has $\cal T$ symmetry only. From condition (2), we find $\sigma^{y}_{xy} = 0$ for $H_A$ and $\sigma^{x}_{xy} = 0$ for $H_B$.

(b). MPG 3. For $H_{25} = R^-_{IP} + R^+_{IP} + D^+_{IP} + D^-_{OP}$ with 3, direct calculation gives $c_x \sim k_x [-2(k_x+k_y)a_3 -3b_2 (-k_x +b_2 + a_2)]$ and $c_y = k_x [(1-2k_y +2k_x) a_3 + 3b_2 (k_y +a_2 +b_2)]$, both $c_{x/y}$ contain a term $\cos3\theta$. Since $d^2 = A+ B \sin3\theta + C \cos3\theta +D \cos6\theta$ (if $R^-_{IP}$ is turned off, $d^2 = A+ D \cos6\theta$), we conclude that $\sigma^{x/y}_{xy} \ne 0$. Switching off $R^-_{IP}$ will make $\sigma^{x/y}_{xy} = 0$ according to condition (5). In addition, $H = H_{4x} + H_{4y} + H_{13y}= R^+_{IP} + D^+_{IP} + R^-_{OP}$ also has $C_3$ symmetry. From condition (5), $\sigma^{x/y}_{xy} = 0$.

(c). MPG 31'. For $H_{31} = R^-_{IP} + R^-_{OP} + D^-_{OP}$ with 31', direct calculation gives $c_x \sim k_x^2 (b_2 -a_2) \sim 1 + \sin2\theta + \sin4\theta + \cos2\theta + \cos4\theta$ and $c_y = k_x (a_3 +b_3) - 3k_x k_y (b_2- a_2) \sim 1 + \sin2\theta + \sin4\theta + \cos2\theta + \cos4\theta$. Thus both $\sigma^{x/y}_{xy} \ne 0$. Note that the system with 31' symmetry can have SOI Hamiltonians different from $H_{25}$. For instance, both $H_A = H_8 +H_{13x}$ and $H_B = H_8 +H_{13y}$ also have 31' symmetry. From condition (2), we find $\sigma^{y}_{xy} = 0$ for $H_A = D^-_{IP} + D^-_{OP}$ and $\sigma^{x}_{xy} = 0$ for $H_B = D^-_{IP} + R^-_{OP}$.

\medskip
\noindent{\bf MPG m', 3m, 3m'} --- For these three MPGs with m=$m_x$, $\sigma^{x} _{xy} = 0$ and $\sigma^{y}_{xy} = 0/1$ according to the notation of Table \ref{table3}. Now we analyze these MPGs in details.

(a). MPG m'. For $H_{20x} = R^-_{IP} + R^+_{IP} + D^+_{OP}$ with $m_x$', we have $\sigma^x_{xy}=0$ from condition (4). Direct calculation gives $c_y \sim 1 + \cos(\theta) + \cos(2\theta) + \cos(3\theta) + \cos(4\theta)$ and $d^2 = A + B \sin(3\theta) + C \cos(4\theta)$. Hence $\sigma^y_{xy} \ne 0$. Note that $H = H_{3y} + H_{13x} = R^+_{IP} + D^-_{OP}$ and therefore $\sigma^{x/y}_{xy}=0$ from condition (2). For $H_{20y} = R^-_{IP} + D^+_{IP} + D^+_{OP}$ with $m_y$', Since $d^2$ has $M_x$ symmetry, we have $\sigma^y_{xy}=0$ from condition (3). $c_x \sim 1 + \sin(\theta) + \sin(3\theta) + \sin(4\theta) + \cos(4\theta)$ and $d^2 = A + B \cos\theta + C \cos(4\theta)$. Hence $\sigma^x_{xy} \ne 0$. Similarly, $H = H_{3x} + H_{13y}$ with $m_y$' makes $\sigma^x_{xy} \ne 0$.

(b). MPG 3m. For $H_{32x} = R^-_{IP} + D^+_{IP} + D^-_{OP}$ with 3$m_x$, we have $\sigma^{x}_{xy} = 0$ from condition (3). Direct calculation gives $c_y = k_x[(1-2k_y)a_3 - 3b_2 (k_y +\cos2\theta)] \sim 1+ \sin3\theta + \sin\theta + \sin5\theta + \cos2\theta +\cos4\theta$ making a nonzero contribution to $\sigma^{x}_{xy}$ since $d^2 = A + B \sin3\theta + C \cos6\theta$. For $H_{32y} = R^-_{IP} + R^+_{IP} + R^-_{OP}$ with 3$m_y$, we have $\sigma^{y}_{xy} = 0$ from condition (4). Direct calculation yields $c_x =k_x [-2k_y b_3 +3a_2 (-k_x +a_2)] \sim 1 + \cos\theta + \cos2\theta +... $ and hence $\sigma^{x}_{xy} \ne 0$.

(c). MPG 3m'. For $H_{24x} = R^-_{IP} + R^+_{IP} + D^+_{OP}$ with 3$m_x$', we have $\sigma^x_{xy}=0$ from condition (4). Direct calculation yields $c_y = k_x[(1+2k_x) a_56 - 6b_5 (k_y +\sin2\theta)$ (which contains $\cos6\theta$) and $d^2 \sim A +B \cos3\theta + C \cos6\theta$ (after expansion, it contains $\cos6\theta$ as well). We conclude that $\sigma^y_{xy} \ne 0$. For $H_{24y} = R^-_{IP} + D^+_{IP} + D^+_{OP}$ with 3$m_y$', we have $\sigma^y_{xy} = 0$ from condition (3) since $d^2$ has $m_x$ symmetry. Direct calculation gives $c_x = k_x(-2k_y a_6 +6 b_5 (k_x+\sin2\theta))$ which contains $\sin3\theta$ and $\cos6\theta$. Note that $d^2 \sim A +B \sin3\theta + C \cos6\theta$, thus $\sigma^x_{xy} \ne 0$. Therefore $\sigma^{x/y}_{xy} = 0/1$.

\medskip
\noindent{\bf For the rest of MPGs} --- For the following SOI Hamiltonians: $H_{11} = R^-_{IP} + D^+_{OP}$ with 4'm'm, $H_{12} = R^-_{IP} + R^+_{OP}$ with 4'm'm, and $H_{14} = R^-_{IP} + D^+_{OP}$ with 4m'm', $H_{15} = R^-_{IP} + R^+_{OP}$ with 4mm, $H_{16} = R^-_{IP} + D^+_{OP}$ with 6m'm', $H_{17} = R^-_{IP} + R^+_{OP}$ with 6mm, $H_{21} = D^-_{IP} + R^+_{OP}$ with MPG 2, $H_{22} = R^-_{IP} + D^+_{OP} + D^+_{OP}= R^-_{IP} + D^+_{OP}$ with 2m'm', $H_{29} = R^-_{IP} + R^+_{OP} + R^+_{OP} = R^-_{IP} + R^+_{OP}$ with 2mm, $H_{30} = R^-_{IP} + R^+_{OP} + D^+_{OP}$ with 4', we have $\sigma^{x/y}_{xy}=0$ from condition (2). For $H_{23} = R^-_{IP} + D^-_{IP} + D^+_{OP}$ with MPG 4 and $H_{26} = R^-_{IP} + D^+_{OP} + R^+_{OP}$ with MPG 6, $\sigma^{x/y}_{xy} = 0$ from condition (5).


\begin{thebibliography}{00}
\bibitem{suppress}
The suppression of linear Rashba SOI was discussed in Ref.~[\onlinecite{Shanavas}]. A transition from the cubic Rashba effect to the coexistence of linear and cubic Rashba effects was observed experimentally in oxide heterostructures\cite{Lin}. For heavy holes in III-V semiconductor quantum wells, the linear Rashba SOI can be absent making the cubic SOI as the leading order\cite{Schliemann,Winkler,Nomura}.
\end{thebibliography}
\end{document}